\documentclass[superscriptaddress,amsmath,amssymb,nofootinbib]{revtex4}
\RequirePackage{lineno}
\usepackage{graphicx}
\usepackage{dcolumn}
\usepackage{bm}
\usepackage{ulem}
\usepackage{xfrac}

\usepackage[printwatermark]{xwatermark}
\usepackage{xcolor}

\usepackage{graphicx,amsmath,amssymb}
\usepackage{lineno}
\usepackage[colorlinks=true, urlcolor=blue, linkcolor=blue, citecolor=blue]{hyperref}
\usepackage{lipsum}

\usepackage{mathtools}

\begin{document}

\title{Studying nuclear medium modification using the Gerasimov--Drell--Hearn sum rule}

\author{A.~Deur} 
\affiliation{Thomas Jefferson National Accelerator Facility, Newport News, Virginia 23606, USA}
\author{M.~M.~Dalton} 
\affiliation{Thomas Jefferson National Accelerator Facility, Newport News, Virginia 23606, USA}
\author{S.~\v{S}irca} 
\affiliation{Faculty of Mathematics and Physics, University of Ljubljana, 1000 Ljubljana, Slovenia}
\affiliation{Jo\v{z}ef Stefan Institute, 1000 Ljubljana, Slovenia}

\begin{abstract}

The Gerasimov--Drell--Hearn sum rule is a generic relation that has been used to make significant contributions to research in hadronic physics. It connects 
the spin-dependent cross-section for photoproduction off a particle to the squared ratio 
of the particle's anomalous magnetic moment and its mass, $(\kappa/M)^2$. Thus, for a nucleon
embedded in a nucleus, the sum rule relates the cross-section to $\kappa/M$ averaged quadratically 
over the nucleons comprising the nucleus.  This quadratic averaging can be used to constrain 
the mechanism responsible for the medium modification of the nucleon.
We also point out that the global properties of the embedded nucleon like its axial charge, mass or magnetic moment are observables measurable through sum rules.  

\end{abstract}

\maketitle

\section{Introduction} 

 As it was first observed by the EMC experiment at CERN \cite{EuropeanMuon:1983wih}, the structure functions of nucleons embedded in nuclei differ from those of free nucleons.  This reveals that the surrounding nuclear medium affects the distribution of quarks and gluons within the nucleon.  The finding was surprising because the four-momentum transfer characterizing the reaction studied by the EMC experiment---deep inelastic scattering (DIS) of leptons off nuclei---is much larger than the scale characterizing nuclear binding: GeV as opposed to MeV, respectively.
 The finding became known as the EMC effect. While understanding it is crucial to reach a comprehensive understanding of hadronic and nuclear structures \cite{Cloet:2019mql}, no consensus on its origin exists \cite{Thomas:2018kcx}. There are currently two leading scenarios: the mean field (MF) mechanism \cite{CiofidegliAtti:2015lcu} and the short-range correlation (SRC) picture \cite{Frankfurt:1988nt,CLAS:2019vsb,Hen:2013oha}. In the MF explanation, all nucleons are similarly modified by their nuclear environment, while in the SRC framework, about $20\,\%$ of the nucleons are involved in SRC pairs and only those are modified. In the SRC case, the nuclear modifications on an affected nucleon are thus about five times larger compared to the MF scenario, but overall they average to the same global nuclear effect that is experimentally determined. 
 
In this article, we show that the value of the Gerasimov--Drell--Hearn (GDH) sum rule~\cite{Gerasimov:1965et} for an embedded nucleon depends on how uniformly the average modification is spread over all the nucleons, with the MF (uniform) and SRC (highly non-uniform) scenarios as examples.
We first describe the GDH sum rule and show how it can disentangle between these scenarios. Next, we quantitatively illustrate 
this point by using these two scenarios and another possibility examined in~\cite{Thomas:2018kcx}. We then discuss practical matters for possible experimental investigations and conclude.
We assume, as is customary in nuclear models, that while embedded nucleons may have modified properties, they preserve their individuality, meaning that their wave-functions do not significantly overlap.

\section{The Gerasimov--Drell--Hearn sum rule} 
The GDH sum rule \cite{Gerasimov:1965et,Helbing:2006zp, Deur:2018roz} is a general relation of quantum field theory that links the anomalous magnetic moment $\kappa$ of a particle and its mass $M$ to the spin-dependent photoproduction cross-section on that particle. This relation stems from causality, unitarity, and Lorentz and gauge invariances. 
The validity of the sum rule rests on the expectation that the cross-section drops off rapidly enough at very high energies. In the case of a hadronic target, this is consistent with both Regge theory \cite{Regge-theory} and current experimental knowledge~\cite{Helbing:2006zp, Dutz:2003mm,Adhikari:2017wox, Zheng:2021yrn, Sulkosky:2019zmn, CLAS:2024fcf}. 
For a nucleon, the sum rule is
\begin{equation}
I \equiv \int_{\nu_0}^{\infty}\frac{\Delta \sigma(\nu)}{\nu}\,\mathrm{d}\nu=\frac{2\pi^2 \alpha \kappa^2}{M^2} \>,
\label{eq:gdh}
\end{equation}
where $\nu$ is the incoming photon energy, $\nu_0$ is the photoproduction energy threshold, 
$\alpha$ is the fine-structure constant, and $\Delta \sigma \equiv \sigma_{P} - \sigma_{A}$ is the difference of total photoproduction cross-sections for which the photon spin is parallel and anti-parallel to the nucleon spin, respectively.
The right-hand side of the sum rule is referred to as its static part, 
$I_{\rm stat} \equiv 2\pi^2 \alpha(\kappa/M)^2$,
and the left-hand side  as the integral, or dynamical, part of the sum rule,
${I_{\rm int}} \equiv \int \Delta \sigma (\mathrm{d}\nu/\nu)$.
We assume that the GDH sum rule is valid for a free nucleon, which is expected from the fundamental nature of the assumptions on which the sum rule derivation rests, along with Regge theory and experimental data.

In the next Section, we show how the MF and SRC scenarios lead to distinct values of $I$ for an embedded nucleon. Measuring one of the sides of the sum rule, presumably the integral part as it is the most feasible, is then sufficient to discriminate MF from SRC if the sum rule value can be calculated within these scenarios. This is not possible with the nonperturbative methods presently available to calculate the Strong Force,
but the measurement would test model-dependent assessments of the GDH value as in, say, \cite{Bass:2020bkl}.
A model-independent test requires one to measure $\Delta \sigma$ and $\kappa/M$ independently. We discuss this in Sect.~\ref{sec:kappa and M}, but first we show that the presence of the squared term in $I_{\rm stat}$
---while there is none in $I_{\rm int}$
---leads to MF and SRC predicting distinct values of the GDH components. In other words, if MF and SRC predict the same $\kappa^*/M^*$, they must predict different $\Delta \sigma^*$. (Throughout this paper, asterisks denote quantities modified by nuclear medium, averaged over all nucleons.) Conversely, with the same $\Delta \sigma^*$ value, they must produce distinct $\kappa^*/M^*$ ratios for the GDH sum rule to be valid. This feature, which is the main point of our paper, offers a way to test the validity of medium modification scenarios by measuring the GDH sum rule. 
 

\subsection{Medium modification of the sum rule \label{modification of GDH}}
We quantify by $\zeta$ the nuclear medium modification  of $I_{\rm int}$,
\begin{equation}
{I^*_{\rm int}}= \int \Delta \sigma^* \frac{ \mathrm{d}\nu}{\nu}   \equiv   \zeta
  \int   \Delta \sigma \frac{ \mathrm{d}\nu}{\nu} =   \zeta I_\mathrm{int}\>.
  \label{Istarint}
\end{equation}
Similarly, let $\chi$ stand for the modification of $\kappa/M$ by nuclear medium,
\begin{equation}
\frac{\kappa^*}{M^*} \equiv  \chi \frac{\kappa}{M}. \nonumber
\end{equation}
Naively, one would expect  
\begin{equation}
\frac{I^*_{\rm stat}}{2\pi^2 \alpha} =  \left(  \chi
 \frac{\kappa}{M}\right)^2 \equiv \left(\frac{\kappa^*}{M^*}\right)^2  =  \chi^2 \frac{I_\mathrm{stat}}{2\pi^2\alpha} \>, \nonumber
\end{equation}
but we will see that $I^*_{\rm stat} = \chi^2 I_{\rm stat}$ actually holds true only for uniform modification scenarios, see Eq.~(\ref{Eq. MF}).
Crucially, $I^*_{\rm stat} \neq \chi^2 I_{\rm stat}$ for non-uniform scenarios, see Eq.~(\ref{Eq. GDH SRC}). 
In other words, $ \chi ^2$ does not necessarily quantify the modification of the static part of GDH, depending on the mechanism altering the ratio $\kappa/M$ of the embedded nucleon.

\subsubsection{Variation in GDH expectations for different  medium modification scenarios \label{GDH-medmod}}
We assume that in the MF scenario, all nucleons are modified equally by the nuclear medium, so if 
$ \chi$ is the average modification of $\kappa/M$ occurring in the nucleus, statistically each nucleon is  modified by the same factor,
\begin{equation}
 \chi =  \chi_\mathrm{MF} \>.
\label{Eq. MF zeta-star}
\end{equation}
Together with Eq.~(\ref{Istarint}) one has
\begin{equation}
I^*_\mathrm{MF} 
  =   \zeta_\mathrm{MF} I_{\rm int} 
  =  \chi_\mathrm{MF}^2 I_{\rm stat} \>.
\label{Eq. MF}
\end{equation}
In the SRC scenario, about $20\,\%$ of nucleons are modified by the nuclear medium \cite{Hen:2013oha}, while the remaining nucleons are unmodified. Since the average modification is $\chi$, the $\kappa/M$ ratio is modified as follows (assuming exactly $20\,\%$ of equally modified nucleons for simplicity):
\begin{equation}
\frac{\kappa^*}{M^*} = 0.2 \,  \frac{\kappa^{*\rm(SRC)}}{M^{*\rm(SRC)}} + 0.8 \frac{\kappa}{M} 
\equiv  0.2 \,  \chi _{\rm SRC}\frac{\kappa}{M} + 0.8 \frac{\kappa}{M} 
\equiv \chi\frac{\kappa}{M} \>,
\label{Eq. SRC kappa/m}
\end{equation}
where $\kappa^{*\rm(SRC)}/M^{*\rm(SRC)}$ is the nuclear modification on nucleons in SRC pairs (in contrast to
$\kappa^{*}/M^{*}$ that is the average modification over all nucleons composing the nucleus).
Equation~(\ref{Eq. SRC kappa/m}) then implies
\begin{equation}
 \chi = 0.2 \,  \chi _{\rm SRC} + 0.8 \>.
\label{Eq. SRC chi-star}
\end{equation}
Likewise, $I_\mathrm{int}$ becomes $(0.2\,  \zeta_{\rm SRC}+0.8) I_{\rm int}$.
 
To determine the corresponding modification to the static part, consider how the GDH relation 
is actually studied. The cross-section difference can be written as
\begin{equation}
\Delta \sigma =\frac{1}{P_\gamma P_T P_N \rm (\gamma\mbox{-}flux) (target~thickness) (efficiencies)}\sum (E^+ - E^-)  \>,
\label{Eq. delta sigma}
\end{equation}
where $\gamma$-flux is the photon flux, $E^\pm$ are the event counts for incoming photons of $\pm$ helicities, respectively, $P_\gamma$ is the photon beam polarization, $P_T$ is the experimental target polarization and $P_{N}$ is the effective nucleon polarization in the target nucleus, which must be obtained theoretically. The summation sign emphasizes that these counts are accumulated. Thus, when the GDH integral is measured, the events originating both from the nucleons involved in SRC pairs and from the unpaired nucleons are summed, corresponding to the $(0.2\,  \zeta_{\rm SRC}+0.8)$ weight for $I_\mathrm{int}$. Assuming for now  $P_{N}=1$ for all nucleons (paired or unpaired), we have
\begin{eqnarray}
\int \Delta \sigma^* \frac{ \mathrm{d}\nu}{\nu} 
&=& 0.2 \,  \zeta_{\rm SRC} \int  \Delta \sigma \frac{ \mathrm{d}\nu}{\nu}  
   +0.8 \int\Delta \sigma \frac{\mathrm{d}\nu}{\nu} \nonumber\\ 
&=& 0.2 \left[2\pi^2 \alpha \left( \frac{\kappa^{*\rm(SRC)}}{M^{*\rm(SRC)}}\right)^2 \right] 
   +0.8 \left[2\pi^2 \alpha \left( \frac{\kappa}{M}\right)^2 \right]  
 =\left( 0.2 \,  \chi_{\rm SRC}^2 + 0.8 \right)I_{\rm stat} \>.
   \label{Eq. delta sigma SRC} 
\end{eqnarray}
The second line is obtained by applying the sum rule, which is valid for any type of particle, be it an unpaired or SRC-paired nucleon.\footnote{Nucleons have Fermi momentum. That associated with the SRC term in Eq.~(\ref{Eq. delta sigma SRC}) can be large, which may modify $I_{\rm int}$ since the sum rule assumes an object at rest. We neglect this possibility since the effect enters at 2$^{\rm nd}$ order (an integral is unaffected by the smearing of its integrant) and is correctable, and because high-momentum components are suppressed in the wave-function generally and by the 0.2 factor in Eq.~(\ref{Eq. delta sigma SRC}).}   This shows that in the SRC case 
$\kappa/M$ must be quadratically averaged for the GDH sum rule to be valid, rather than squaring the average $(0.2\,{ \chi_{\rm SRC}} + 0.8)$, yielding the relation
\begin{equation}
I_\mathrm{SRC}^* 
  = (0.2\,  \zeta_{\rm SRC}+0.8)I_{\rm int} 
  = \left(0.2\, \chi_{\rm SRC}^2 + 0.8 \right) I_{\rm stat} \>.
  \label{Eq. GDH SRC}
\end{equation}
This contrasts with the uniform modification case, $I^*_{\rm stat} = 2\pi^2 \alpha \left(  \chi
 \kappa/M\right)^2$, which, together with Eq.~(\ref{Eq. SRC chi-star}), would yield an incorrect relation $( 0.2\,  \zeta_{\rm SRC}+0.8)I_{\rm int} 
  = \left(0.2\, \chi_{\rm SRC} + 0.8 \right)^2 I_{\rm stat}$.

Thus, from Eqs.~(\ref{Eq. MF zeta-star}) and (\ref{Eq. MF}) the sum rule in the MF scenario is
\begin{equation}
I_\mathrm{MF}^*={ \chi}^2 I \>,  \nonumber
\end{equation}
while for the SRC case Eqs.~(\ref{Eq. SRC chi-star}) and (\ref{Eq. GDH SRC}) yield
\begin{eqnarray}
I_\mathrm{SRC}^*= \left( \frac{( \chi-0.8)^2}{0.2} +0.8  \right) I \>. \nonumber
\end{eqnarray}
This insight is the central point of the article.

A refinement of the SRC scenario is to consider the polarization of the embedded nucleons. Pairs of nucleons undergoing SRC  will be scattered into a high relative momentum $D$-wave state by the nuclear tensor force. Ref.~\cite{Thomas:2018kcx} pointed out that a SRC nucleon should experience ``$D$-wave depolarization'' (DWD), where the two units of angular momentum of the $D$-wave come from the reversal of the polarized nucleon spin direction. This should lead to an effective polarization of about $-0.15 \lesssim P_{D} \lesssim -0.1$ for nucleons in the $D$-wave. 
Following the previous argument and accounting for different nucleon polarizations,  Eq.~(\ref{Eq. delta sigma SRC}) is modified to 
\begin{eqnarray}
\int \Delta \sigma^* \frac{ \mathrm{d}\nu}{\nu} 
&=& \frac{1}{P_{N}} \left( 0.2 \, P_{D} \zeta_{\rm SRC} \int  \Delta \sigma \frac{ \mathrm{d}\nu}{\nu}  
   +0.8 P_{S} \int\Delta \sigma \frac{\mathrm{d}\nu}{\nu} \right) \nonumber\\ 
&=& \frac{1}{P_{N}} \left( {0.2}{P_{D}} \left[2\pi^2 \alpha \left( \frac{\kappa^{*\rm(SRC)}}{M^{*\rm(SRC)}}\right)^2 \right] 
   + {0.8}{P_{S}} \left[2\pi^2 \alpha \left( \frac{\kappa}{M}\right)^2 \right]  \right)
 = \frac{1}{P_{N}} \left( {0.2}{P_{D}}\chi_{\rm SRC}^2 + {0.8}{P_{S}} \right)I_{\rm stat} \>, \label{Eq. delta sigma SRC DWD} 
\end{eqnarray}
where 
$P_{S}$ is the $S$-wave polarization
and $P_{N}={0.2}P_{D} + 0.8P_{S}$ makes explicit the correction for the average nuclear polarization from Eq.~(\ref{Eq. delta sigma}). 
Eq.~(\ref{Eq. delta sigma SRC DWD} ) clarifies that  polarization techniques are sensitive to modifications only proportional, in sign and magnitude, to the polarization of the modified nucleons.  We then obtain, analogous to Eq.~(\ref{Eq. GDH SRC}),
\begin{equation}
I_\mathrm{DWD}^* 
  = \frac{1}{P_{N}} \left( {0.2}\, P_{D}  \zeta_{\rm SRC} + 0.8 P_{S} \right)I_{\rm int} 
  = \frac{1}{P_{N}} \left( {0.2}\, P_{D} \, \chi_{\rm SRC}^2 + 0.8  P_{S}\right) I_{\rm stat} \>.
  \label{Eq. GDH SRC DWD}
\end{equation}
Using Eqs.~(\ref{Eq. SRC chi-star}) and (\ref{Eq. GDH SRC DWD}) we obtain 
\begin{equation}
    I_\mathrm{DWD}^*   = \frac{1}{P_N}  \left( P_{D}\frac{ (\chi - 0.8P_{S})^2}{0.2} + 0.8P_{S}  \right) I. \nonumber
    \end{equation}
Using $P_{D} = -0.1$ and $P_{S} = +1$, which yields $P_{N} = 0.78$, we get
\begin{equation}
I_\mathrm{DWD}^*  = \frac{1}{0.78} \left( -0.5\ (\chi - 0.8)^2 + 0.8  \right) I. \nonumber
\end{equation}
In this scenario, increasing modification tends to decrease the integral as this modification comes from nucleons which contribute with opposite polarization. 

\begin{figure}[!htbp]
\centering
\includegraphics[width=0.38\textwidth]{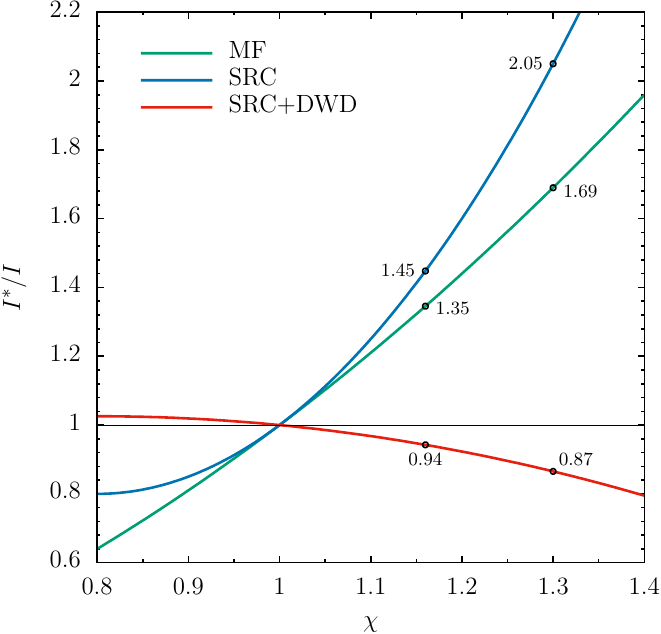}
\caption{Expected value of the medium-modified GDH sum, $I^*$, normalized to the free-nucleon case, $I$, as a function of $ \chi$, the average nuclear modification of $\kappa/M$.  The dependencies for light nuclei ($N\approx Z$) based on the MF, SRC and SRC-DWD scenarios 
are shown by green, blue and red curves, respectively.
The specific values of $I^\ast/I$ at $ \chi=1.16$ and $1.30$ are given next to each curve.}
\label{fig:med.mod}
\end{figure}

It follows that for a given average medium modification $\chi$, the quadratic averaging present in GDH is very sensitive to the uniformity of the modification over the nucleons and to the correlation of the modification with the polarization.
This is shown in Fig.~\ref{fig:med.mod} where, for predicted values of $\chi=1.30$ \cite{Bass:2020bkl} and $\chi=1.16$ \cite{Dalton:2023LOI} (both based on the quark–meson coupling 
model~\cite{Saito:2005rv}), the expectations for the MF, SRC, and SRC+DWD scenarios differ significantly.

Note that the quadratic averaging occurring in the GDH static part is unusual: most sum rules have a static part that depends linearly on the nucleon's static attributes. Thus, they cannot distinguish between medium modification scenarios as described above. 
For example, the unpolarized equivalent of the GDH relation, the Baldin sum rule \cite{baldinSR}, has a static part proportional to $\alpha+\beta$, the sum of electric and magnetic polarizabilities.

\subsubsection{$Z$-dependence of the GDH sum rule for an embedded nucleon  \label{stat_z_dep}}
In practice, the nuclei that are 
most advantageous for realizing a nuclear polarized target are light- to medium-sized.
In the SRC scenario, the probability for a proton to be in a proton--neutron pair increases as $N/Z$, while that of a neutron is approximately  constant.  For $Z\simeq 40$, say, one has $N\simeq 50$, so $\sim$20$\,\%$ of the neutrons and $\sim$25$\,\%$ of the protons are in SRC pairs.   For simplicity, one may assume $ \chi$ to be the same for neutrons and protons, without affecting the conclusion of this discussion. 
Then, for $Z\simeq 40$ nuclei in the SRC case, Eq.~(\ref{Eq. GDH SRC}) remains unchanged for the neutron, while for the proton it becomes ${I^{p*}_{\rm stat}} = (0.25\, \chi_{\rm SRC}^2 + 0.75) I^p_{\rm stat}$, where $ \chi_{\rm SRC}$ is the same as in Eq.~(\ref{Eq. GDH SRC}), independently of $(A,Z)$, as shown by SRC data \cite{Frankfurt:1988nt,CLAS:2019vsb,Hen:2013oha}.
Since the $Z$-dependence enters linearly (not quadratically) in front of ${ \chi}^2$, there is no benefit of studying 
the $Z$-dependence of medium modifications in the context of separating  the MF and SRC scenarios.  In fact, the difference between SRC and MF expectations 
for GDH on the embedded proton decreases for nuclei with $N>Z$. 
For example, assuming that $ \chi_{\rm MF}$ is independent of $Z/N$, the $21\,\%$ enhancement of SRC over MF for $ \chi=1.3$ for light nuclei ($N = Z$) drops to $16\,\%$ for heavier nuclei. (In this discussion, we assume the SRC scenario, not SRC+DWD where the polarization of paired nucleons is different from the average.) This lower sensitivity is expected since the larger the fraction of SRC-paired protons, the closer Eq.~(\ref{Eq. SRC chi-star}) gets to Eq.~(\ref{Eq. MF zeta-star}). In the limit where all protons are in SRC pairs, these two equations coincide and result in the same expectation for the modification of the GDH relation. 

Another disadvantage of heavier nuclei is that there are more scatterings from unpolarized nucleons. Since experiments are typically limited by data acquisition rates, this reduces the statistical significance of the experiment. That is, if an asymmetry is measured to determine $\Delta\sigma^*/\sigma^*_{\rm tot}$, larger nuclei will cause a larger dilution of that asymmetry.

\section{Determination of $\kappa^*$ and $M^*$ \label{sec:kappa and M}}

Although the primary purpose of this article is to show how the GDH sum rule can distinguish between 
medium modification scenarios (Sect.~\ref{GDH-medmod}), we discuss here what such a test would entail in practice---assuming 
that either the MF or the SRC scenario is realized, and not a combination of them.
To be able to actually perform the test, the values of $\kappa^*$ and $M^*$---or at least $\kappa^*/M^*$---are necessary, and 
must be obtained by different means than by a GDH measurement itself.
Then the independently determined $I^*_{\rm int}$ and $ \chi$ (from $\kappa^*/M^*$) values 
can be related 
as in Fig.~\ref{fig:med.mod}, providing a unique perspective on the underlying nature of medium modifications.  
Several strategies are possible to determine $\kappa^*$ and $M^*$, 
and the discussion below is illustrative rather than exhaustive.

\subsection{Separation of $\kappa^*$ and $M^*$ using the GDH and Schwinger sum rules}

The modified anomalous magnetic moment $\kappa^*$ and mass $M^*$ of a charged particle can be separated by using the GDH and Schwinger sum rules in conjunction. The latter sum rule is \cite{Schwinger:1975ti}
\begin{equation}
I_{\rm LT}(Q^2) \equiv \int_{\nu_0}^{\infty} f_\gamma \frac{\sigma_{\rm LT}(\nu,Q^2)}{Q\nu}\,\mathrm{d}\nu \xrightarrow[Q^2 \to 0]{}    \frac{\pi^2 \alpha \kappa e_t}{M^2},
\label{eq:Sch-SR}
\end{equation}
where $Q$ is the virtuality of the photon,  
$f_\gamma$ is the virtual photon equivalent energy~\cite{note on photon flux}, 
$\sigma_{\rm LT}$ is the longitudinal-transverse interference cross-section~\cite{Deur:2018roz} and
$e_t$ is the electric charge of the particle in units of elementary charge. 
For instance, $e_t = 1$ for a proton, regardless of whether it is free or embedded in a nucleus since $e_t$ 
is fixed by charge conservation.  The Schwinger sum rule is studied by inclusive polarized electroproduction at $Q^2\ne 0$, and the data are then extrapolated to the real-photon point, $Q^2=0$. Such a measurement has been performed at very low $Q^2$ for the neutron embedded in $^3$He, with the sum rule being verified after extrapolation \cite{E97-110:2021mxm}.  The Jefferson Lab low-$Q^2$ GDH program
\cite{Adhikari:2017wox, Zheng:2021yrn, Sulkosky:2019zmn, CLAS:2024fcf, E97-110:2021mxm} also demonstrated that measurements of the GDH sum on the proton, neutron, deuteron and $^3$He by using electroproduction at very low $Q^2$ are competitive with direct photoproduction measurements: the two experimental approaches involve systematic uncertainties of different origin but of comparable size. 
Hence, both the GDH and Schwinger sum rules could be measured concomitantly in a low-$Q^2$ electroproduction experiment; this would yield different ratios $\kappa^{*2}/M^{*2}$ and $\kappa^*/M^{*2}$ from which $\kappa^*$ and $M^*$ could be separated by using the relations
$$
\kappa^* = \frac{I^*}{2I^*_{\rm LT}} \>, \quad
M^* = \pi \sqrt{\frac{\alpha}{2}} \frac{\sqrt I^*}{I^*_{\rm LT}} \>, 
$$
where $I^*_{\rm LT}$ denotes the measured Schwinger sum for the medium-modified proton. 

Several important remarks are in order.
The quantities $\kappa^*$ and $M^*$ defined in this manner are obtained from actual measurements and are therefore unambiguous.  Their specific meanings---for example, whether they are the genuine magnetic moment and mass of an embedded proton, or effective quantities---are irrelevant for the separation of the MF and SRC scenarios. 
(This said, measuring these quantities is interesting on its own.)  The above separation works for the embedded proton, not for the neutron whose charge $e_t$ is zero. 

Separating $\kappa^*$ and $M^*$ is a first step toward discriminating model-independently the scenarios 
for medium modifications.
To go further, we need an independent determination of either $\kappa^*$ or $M^*$.  Next we discuss two possibilities.

\subsection{Determination of $\kappa^*$ from magnetic form-factor measurements}
The modified $\kappa^*$ can be obtained from a measurement of the embedded nucleon's 
magnetic moment $\mu^*=e_t+\kappa^*$. For example, $\mu^*$ is the $Q^2=0$ limit of the magnetic form-factor $G^*_M$ of the embedded nucleon, which has been measured \cite{Yakhshiev:2002sr, A1:2020toz}.  
Following Eq.~(11) of \cite{A1:2020toz}
for the nuclear modification of the form-factor ratio $G_E/G_M$,
$$
a_{\rm mod} \equiv (G^*_E/G^*_M)\bigl/(G_E/G_M) \>,
$$
and since $e_t$ is the $Q^2=0$ limit of both $G_E$ and $G^*_E$ due to charge conservation, we have for the proton: 
$$
\kappa_p^*= \mu_p^*-1=\frac{\kappa_p+1}{a_{{\rm mod},p}}-1 \>,
$$
and for the neutron:
$$
\kappa_n^*= \mu_n^*=\frac{\kappa_n}{a_{{\rm mod},n}} \>.
$$
Then, $\kappa^*$ together with Eq.~(\ref{eq:Sch-SR})
and the measured $I^*_{\rm LT}$ determine the value 
$$
\frac{\kappa^*}{M^*} = \sqrt{\frac{\kappa^* I^*_{\rm LT}}{\pi^2 \alpha e_t}} \>,
$$ 
which can be compared to the $I^*$ measured by GDH.  
(In practice, only $\kappa_p$ can be used since Eq.~(\ref{eq:Sch-SR}) 
has no information in the neutron case when $e_t=0$ and $I^*_{\rm LT}=0$.)
Comparing the values $I^*$ and $\kappa^*/M^*$ in the sense of Fig.~\ref{fig:med.mod} then discriminates in a model-independent way the frameworks proposed to explain medium modifications.

\subsection{Determinations of $M^*$}

An alternative to using $\kappa^*$ obtained from form-factors is to determine $M^*$,
the effective mass of the embedded nucleon.  One may be tempted, for instance, 
to follow the lead of mean-field calculations suggesting that at nuclear matter density, 
the nucleon and $\Delta$ masses are decreased, such that
${M^*}/{M} \approx {M_\Delta^*}/{M_\Delta} \approx 0.9$ \cite{Saito:2005rv},
the $\Delta$ mass shift in nuclear matter, $M_\Delta^*-M_\Delta$, being clearly 
established \cite{Delta mass shift}.  Yet $M^*$ is often viewed as not directly measurable but a conventional concept characterizing the properties of a quasi-particle within a strongly interacting medium. 
In this view, it is a model-dependent quantity that can only be extracted indirectly
through nuclear models and observables that depend on how nucleons move in the nuclear 
medium. Sum rules, however, illuminate this notion rather differently by implying that global properties of embedded nucleons are indeed observable (Sect.~\ref{sec:emb_ppt_obs}).

Many definitions of 
$M^*$ exist depending on the physics 
context, but the two most frequently used are the non-relativistic mass 
and the relativistic Dirac mass. The former is commonly introduced in non-relativistic 
shell-model calculations with complex single-nucleon potentials in order to render 
these potentials effectively independent
of energy and momenta \cite{Mahaux:1985zz,Hodgson83}; in this sense
the non-relativistic effective mass parameterizes the momentum and energy dependence 
of the potential and hence is a measure of its non-locality 
either in space (momentum dependence) or time (energy dependence), or both.  
The two dependencies can be decoupled, resulting in the 
concepts of ``momentum mass'' and ``energy mass'', but generally the only meaningful
quantity is the product of the two \cite{Li:2018lpy,Jeukenne:1976uy}.
Data indicate that in heavy nuclei, the nucleon mass is generally reduced
to $\approx 70\,\%$ or $80\,\%$ of its free value, and the range of studies 
from which such values of $M^*$ are extracted is broad  \cite{Blaizot81}:  
from the analysis of the density of single-particle states around 
the Fermi surface \cite{BrownNP63}, studying the optical potential 
for proton scattering off nuclei \cite{Jeukenne:1976uy}, the analysis 
of the energy dependence of the imaginary part of the single-particle potential \cite{Mahaux81}, to studies of giant resonances in heavy nuclei \cite{Zhang21,ZhangPRC16,Su20}.

The relativistic framework also provides several interpretations of $M^*$, 
each with distinct motivations and serving specific purposes \cite{Jaminon:1989wj}.
In the most frequent form of the relativistic Dirac mass, medium effects are taken into
account by the scalar and vector parts of the nucleon self-energy, which are used 
as corrections to the bare mass to yield a $M^*$ that can be used, say, 
in relativistic Bruckner \cite{vanDalen:2005ns} or 
Hartree--Fock 
calculations \cite{Li:2015ank}.  The density dependence of $M^*$ and 
its isospin splitting (as a function of density) have also been 
calculated \cite{Zhang:2015qdp,Li:2015pma}.
 
Regardless of the type of the approach from the above list one invariably
finds $M^*/M \approx 0.8 \pm 0.1$, and while it is not the purpose
of this paper to resolve this matter, it remains to be explored whether 
a genuine connection exists between the various theoretically devised $M^*$ 
and the $M^*$ measured by the GDH sum rule; at any rate it would be interesting 
to establish whether the GDH sum rule applied to the embedded nucleon will be 
satisfied with $M^*$ equal to the bare mass or different from it. 

\subsection{Connection to the EMC effect}
 We now explain how GDH studies on an embedded nucleon is directly relevant to studying the EMC effect~\cite{EuropeanMuon:1983wih,Cloet:2019mql,Thomas:2018kcx}. This may initially not seem obvious since the GDH sum rule stands at $Q^2=0$ and spans nearly all energies, $\nu \geq \nu_0$. In contrast, the EMC effect occurs in the DIS domain where $Q^2 >1$~GeV$^2$ and $\nu \gg \nu_0$. 
However, the root cause of the EMC effect need not be restricted to the DIS domain and, furthermore, the GDH sum rule naturally generalizes to non-zero $Q^2$~\cite{Deur:2018roz, Anselmino:1988hn, Ji:1999mr, Drechsel:2000ct}. This generalization essentially replaces $\Delta \sigma(\nu)$ by the first spin structure function of the nucleon, $g_1(\nu,Q^2)$, for which the effects of nuclear matter have been theoretically studied~\cite{Cloet:2005rt}. Crucially, we expect SRC effects to persist at  $Q^2=0$~\cite{GlueX:2020dvv, Sharp:2025bgk}, ensuring that the discriminatory power of GDH is relevant to the EMC effect. 
In fact, it is reasonable to expect that most of the modification of the GDH sum for an embedded nucleon can be attributed to what causes the EMC effect. 
One traditionally distinguishes four phenomena that modify the structure functions of an embedded nucleon: 
(1) Fermi motion, 
(2) anti-shadowing, 
(3) shadowing and 
(4) the underlying cause of the EMC effect. 
Fermi motion has little influence since the GDH sum rule is an integral property, rather insensitive to the smearing of the integrand.
Anti-shadowing and shadowing occur at larger $\nu$ values than the EMC effect 
and are therefore suppressed relative to it by the $1/\nu$ weight in Eq.~(\ref{eq:gdh}). 
The first three types of medium modifications
are expected to be uniform across all nucleons, much like the MF scenario, and any non-uniformity observed using the GDH relation would be associated with the EMC region. 
In all, one expects the GDH value to be predominantly modified by the medium effects that cause, in particular, the EMC effect. Furthermore, the discriminatory power of  GDH applies to these particular medium modifications, while being transparent to other known types of medium effects. Thus, studying  GDH  for an embedded nucleon is directly connected to studying the EMC effect. 

\subsection{Global properties of embedded nucleons as observables}
\label{sec:emb_ppt_obs}
Our proposal to use sum rules to study medium modification 
provides perspectives on the nature of the global properties of embedded nucleons such as $M^*$ or $\kappa^*$. 
A conservative view is that such quantities are not observable and must be interpreted within a given model, which introduces subjectivity and reduces their relevance. In this view, the only genuine observables are what is measured, that is, the 
medium-modified structure functions. 
However, the very definition of a sum rule makes it a relation between structure functions (in the general sense, by including form-factors and structure functions) and global nucleon properties. Thus, if modified structure functions are observables, then so should be the global properties involved in the sum rules. 
This important conclusion relies on two conditions. 
First, the sum rule remains valid for an embedded nucleon. 
This is clearly the case for GDH given the standard 
assumption that nucleons preserve their individuality, without which the whole question of whether quantities such as $M^*$ are observables is moot. This is because GDH is a general relation within quantum field theory, based on causality, unitarity and gauge and Lorentz invariance, generic properties with universal physical validity. The other necessary ingredient of the GDH sum rule is a well-behaved cross-section at large $\nu$. This cannot be altered by nuclear 
matter either; otherwise, the GDH integral on the nucleus containing the nucleon would 
also diverge.  The second condition is that the sum rule for an embedded nucleon 
is measurable. This means that one can separate the nuclear break-up/quasi-elastic
contribution from the reactions at higher energies.  A sum rule including 
the break-up/quasi-elastic part applies to the nucleus, while it applies 
to the (embedded) nucleon if these reactions are excluded. Such a separation is possible 
and has been performed previously  to obtain the sum rules for the neutron
from deuteron or polarized $^3$He data: 
see, for example, \cite{CLAS:2021apd, JeffersonLabE97-110:2019fsc}.
Under these assumptions, global properties of embedded nucleons may be understood as observable through the use of sum rules.

\section{Conclusion}
We have shown that different medium-modification scenarios leading to the same prediction for the EMC effect will result in different expectations for the GDH sum rule for a nucleon embedded in a nucleus. The reason is the  squared averaging that occurs in the static part of the sum rule.  In a uniform modification scenario, medium effects rescale the GDH sum rule by ${ \chi}^2$, with $ \chi$ the relative medium modification of $\kappa/M$. In a non-uniform scenario this could be very different:  
a short-range correlation scenario rescales GDH by $[ \chi-(1-\alpha)]^2/\alpha + 1-\alpha$, with $\alpha$ the fraction of the nucleons in SRC pairs.  
If $D$-wave depolarization is accounted for, then GDH may even decrease.  In practice this would at least depend on how the modification and the effective polarization vary with the relative momentum of paired nucleons.
The discriminatory power of GDH that we describe in this article is important since determining the origin of the EMC effect in particular, and more generally of medium modification, is a central issue in contemporary nuclear and hadronic research.
Ideally, calculations based on any given model would provide specific expectations, which can then be tested by GDH measurements.  One may, however, test model-independently scenarios by supplementing the GDH evaluation by other measurements that provide $\kappa^*$ and $M^*$. 
Finally, we point out that due to the link that sum rules provide between structure functions and global properties of the target, the global properties of embedded nucleons such as their mass, magnetic moment, axial charge etc.~ may be considered as observables measurable through sum rules.

~

\noindent { \bf Acknowledgments}
This material is based upon work supported by the U.S. Department of Energy, 
Office of Science, Office of Nuclear Physics under contracts DE--AC05--06OR23177 
(A.~D., M.~M.~D.) and the Slovenian Research Agency, Core Funding No.~P1--0102 (S.~\v{S}).


\begin{references}
\bibitem{note on photon flux}
While the product $f_\gamma  \sigma_{\rm LT}$ is observable, the individual quantities depends on the
convention chosen for the virtual photon equivalent energy. In the Hand convention~\cite{Hand:1963bb}, it is
$f_\gamma = \nu - Q^2/(2M)$. Another common convention is the Gilman convention for which 
$f_\gamma = \sqrt{\nu^2 +Q^2}$~\cite{Gilman:1967sn}
\end{references}
\end{document}